# МНОГОКОМПОНЕНТНАЯ МОДЕЛЬ ОБУЧЕНИЯ И ЕЕ ИСПОЛЬЗОВАНИЕ ДЛЯ ИССЛЕДОВАНИЯ ДИДАКТИЧЕСКИХ СИСТЕМ

Майер Р.В.

*ФБГОУ ВПО «Глазовский государственный педагогический институт им. В.Г. Короленко», Глазов, email: robert_maier@mail.ru*

При компьютерном моделировании процесса обучения обычно предполагается, что все элементы учебного материала усваиваются одинаково прочно. Но на практике те знания, которые включены в учебную деятельность ученика, запоминаются значительно прочнее, чем знания, которые он не использует. С целью более точного исследования дидактических систем предложена многокомпонентная модель обучения, учитывающая: 1) переход непрочных знаний в прочные; 2) различие в скорости забывания прочных и непрочных знаний. Предполагается, что скорость увеличения знаний ученика пропорциональна: 1) разности между уровнем требований учителя и количеством усвоенных знаний; 2) количеству усвоенных знаний, возведенному в некоторую степень. Рассмотрены примеры использования многокомпонентной модели для изучения ситуаций, возникающих в процессе обучения, представлены получающиеся графики зависимости уровня знаний учащегося от времени. Предложена обобщенная модель обучения, позволяющая учесть сложность различных элементов учебного материала. Рассмотрена возможность создания обучающей программы для тренировки студентов педвузов.

**Ключевые слова:** дидактика, информационно-кибернетический подход, компьютерное моделирование процесса обучения

# MULTI–COMPONENT MODEL OF LEARNING AND ITS USE FOR RESEARCH DIDACTIC SYSTEM

Mayer R.V.

*FSBEI of HPE «The Glazov Korolenko State Pedagogical Institute», Glazov, email: robert_maier@mail.ru*

In computer simulation of the learning process is usually assumed that all elements of the training material are assimilated equally durable. But in practice, the knowledge, which a student uses in its operations, are remembered much better. For a more precise study of didactic systems the multi–component model of learning are proposed. It takes into account: 1) the transition of «weak» knowledge in «trustworthy» knowledge; 2) the difference in the rate of forgetting the «trustworthy» and «weak» knowledge. It is assumed that the rate of increase of student's knowledge is proportional to: 1) the difference between the level of the requirements of teachers and the number of learned knowledge; 2) the amount of learned knowledge, raised to some power. Examples of the use of a multi–component model for the study of situations in the learning process are considered, the resulting graphs of the student's level of knowledge of the time are presented. A generalized model of learning, which allows to take into account the complexity of the various elements of the educational material are proposed. The possibility of creating a training program for the training of students of pedagogical institutes are considered.

Оптимизация процесса обучения требует не только совершенствования содержания и методики изучения отдельных предметов, но и разработки теоретических основ дидактики с привлечением как гуманитарных (психология), так и точных наук (математика, кибернетика). Среди современных методов исследования педагогических систем особое положение занимают **информационно-кибернетический подход** к анализу учебного процесса, основанный на рассмотрении дидактической системы "учитель – ученик" с точки зрения теории управления, а также методы математического и имитационного (компьютерного) моделирования. Их сущность состоит в том, что реальная педагогическая система заменяется абстрактной моделью – некоторым идеализированным объектом, который ведет себя подобно изучаемой системе. Такой моделью может быть система логических правил, математических уравнений [1, 5–8] или компьютерная программа [2–4, 6], позволяющая провести серию экспериментов при различных параметрах, начальных условиях и внешних воздействиях. Изменяя начальные данные и параметры модели, можно исследовать пути развития системы, определить ее состояние в конце обучения.

Сформулируем **основную задачу имитационного моделирования процесса обучения**: зная параметры учащихся, характеристики используемых методов и учебную программу (распределение учебной информации), определить уровень знаний (или сформированности навыка) у учащихся в конце обучения. Известны дискретные и непрерывные модели, основывающиеся на автоматном подходе и решении дифференциальных уравнений [1, 4, 7]. В некоторых случаях используют мультиагентное моделирование, при котором каждый учащийся заменяется программным агентом, функционирующим независимо от других агентов [3]. Также существуют имитационные модели, использующие сети Петри, генетические алгоритмы, матричное моделирование [1–4].

Перечисленные модели имеют общий недостаток: они не учитывают то, что элементы учебного материала (ЭУМ), усвоенные учеником, не равноправны. Те ЭУМ, которые включены в деятельность ученика, превращаются в прочные знания и забываются медленнее, а те, что не включены – быстрее. В процессе учебной деятельности непрочные знания постепенно становятся прочными. **Проблема** состоит в том, чтобы создать имитационную модель обучения, учитывающую различие в скорости забывания различных ЭУМ и переход непрочных знаний в разряд прочных знаний. Было выдвинуто **предположение**: компьютерная имитация будет более точно соответствовать реальному процессу обучения, если учесть, что:

1) прочность усвоения различных ЭУМ неодинакова, поэтому все ЭУМ следует разделить на несколько категорий;

2) прочные знания забываются существенно медленнее непрочных;

3) непрочные знания при их использовании учащимся постепенно становятся прочными.

### Многокомпонентная модель обучения

Известно, что процесс усвоения и запоминания сообщаемой информации состоит в установлении ассоциативных связей между новыми и имеющимися знаниями. В результате приобретенные знания становятся более прочными и забываются значительно медленнее. Многократное использование знаний приводит к формированию у ученика соответствующих умений и навыков, которые сохраняются длительное время.

Обозначим через $U$ уровень требований, предъявляемый учителем и равный количеству $Z_0$ сообщаемых ЭУМ. Пусть $Z$ – суммарные знания ученика, которые включают в себя знания первой, второй, третьей и четвертой категорий: $Z = Z_1 + Z_2 + Z_3 + Z_4$. При этом $Z_1$ – самые непрочные знания первой категории с высоким коэффициентом забывания $\gamma_1$, а $Z_4$ – самые прочные знания четвертой категории с низким $\gamma_4$ ($\gamma_4 < \gamma_3 < \gamma_2 < \gamma_1$). Коэффициенты усвоения $\alpha_i$ характеризуют быстроту перехода знаний $(i-1)$-й категории в знания $i$-й категории. Предлагаемая четырехкомпонентная модель обучения выражается системой уравнений:

$$dZ_1/dt = k\alpha_1(U-Z)Z^b - k\alpha_2 Z_1 - \gamma_1 Z_1;$$
$$dZ_2/dt = k\alpha_2 Z_1 - k\alpha_3 Z_2 - \gamma_2 Z_2;$$
$$dZ_3/dt = k\alpha_3 Z_2 - k\alpha_4 Z_3 - \gamma_3 Z_3;$$
$$Z = Z_1 + Z_2 + Z_3 + Z_4.$$

Пока происходит обучение ($k = 1$), скорость увеличения непрочных знаний ученика пропорциональна: 1) разности между уровнем требований учителя $U$ и общим уровнем знаний $Z$; 2) количеству уже имеющихся знаний $Z$ в степени $b$. Последнее объясняется тем, что наличие знаний способствует установлению новых ассоциативных связей и запоминанию новой информации. Если прирост знаний ученика существенно меньше их общего количества, то $b = 0$. Когда обучение прекращается ($k = 0$), $Z$ уменьшается за счет забывания. Коэффициент забывания $\gamma = 1/\tau$, где $\tau$ – время, в течение которого количество знаний $i$-й категории уменьшается в $e = 2,72...$ раза. Результат обучения характеризуется суммарным уровнем приобретенных знаний $Z = Z_1 + Z_2 + Z_3 + Z_4$ и коэффициентом прочности $\Pr = (Z_2/4 + Z_3/2 + Z_4)/Z$. Если все приобретенные во время обучения знания непрочные ($Z_1 = Z, Z_2 = Z_3 = Z_4 = 0$), то коэффициент прочности $\Pr = 0$. Надо стремиться к ситуации, когда все приобретенные знания прочные ($Z_4 = Z, Z_1 = Z_2 = Z_3 = 0$), тогда $\Pr = 1$. При длительном изучении одной темы уровень знаний $Z$ увеличивается до $U$, затем происходит повышение доли прочных знаний $Z_4$, растет прочность $\Pr$.

### Использование модели обучения

Проанализируем несколько ситуаций, возникающих в процессе обучения.

1. Учитель проводит три урока, уровень требований $U_i$ в течение $i$-го урока задан ($i = 1, 2, 3$). Проанализируем процесс обучения ученика с помощью четырехкомпонентной модели. Результаты моделирования представлены на рис. 1.1. Видно, что во время обучения общее количество знаний $Z$ ученика растет, часть непрочных знаний становится более прочной. Во время перерывов и после обучения уровень непрочных знаний $Z_1$ быстро уменьшается, а прочные знания $Z_4$ забываются существенно медленнее.

2. Учитель проводит три урока, уровень требований $U(t)$ в течение $i$-го урока растет по закону $U_i = a_i(t - t_{i0}) + b_i$, $i = 1, 2, 3$. Проанализируем процесс обучения с помощью двухкомпонентной модели. Двухкомпонентная модель обучения выражается уравнениями:

$$dZ_1/dt = k\alpha_1(U-Z) - k\alpha_2 Z_1 - \gamma_1 Z_1;$$
$$dZ_2/dt = k\alpha_2 Z_1 - \gamma_2 Z_2;$$
$$Z = Z_1 + Z_2.$$

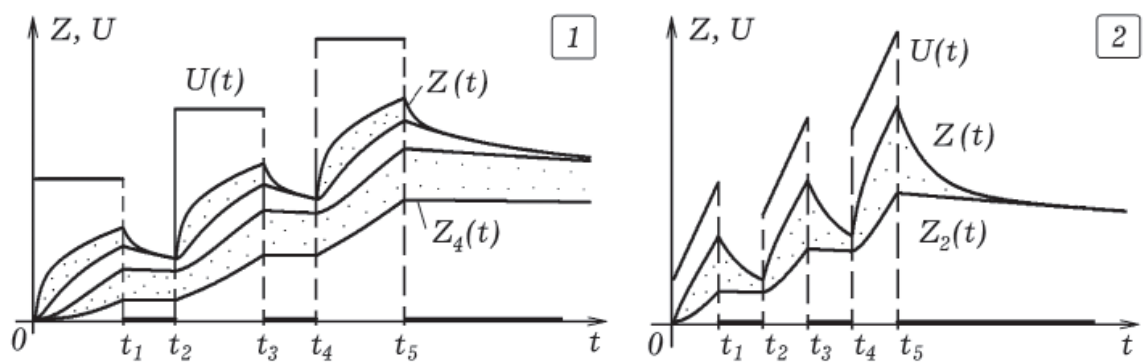

*Рис. 1. Изменение уровня требований U учителя и количества знаний $Z_i$ ученика в процессе обучения*

Результаты моделирования приведены на рис. 1.2. На каждом уроке учитель требует от учащихся:

1) владения материалом, изученным на предыдущих уроках;
2) усвоения новой информации.

Во время обучения непрочные знания становятся прочными и после обучения забываются существенно медленнее.

3. Учитель должен обучить ученика решать $N$ задач возрастающей сложности $\theta_i = i\Delta\theta$, которая считается равной количеству знаний $Z$, требующихся для решения $i$-й задачи. Учитель располагает задачи в порядке возрастания сложности и задает их ученику через равные промежутки времени $\Delta t$. Если ученик не решил $i$-ую задачу, то учитель его обучает в течение времени $\Delta t$, а затем снова предлагает эту же или аналогичную задачу той же сложности $\theta_i$. Если уровень знаний ученика $Z$ больше $\theta_i$, то ученик, вероятнее всего, решит задачу в течение $\Delta t$. При этом $Z$ не увеличится, но часть непрочных знаний станет прочной. После этого учитель предъявляет ему $(i+1+1)$-ую задачу с более высоким уровнем сложности $\theta_{i+1}$. Если у ученика знаний недостаточно, то с большой вероятностью он не сможет решить задачу сразу. Учитель в течение времени $\Delta t$ объясняет материал, либо ученик занимается по учебнику; уровень требований $U = \theta_i$, знания $Z_1$ и $Z_1$ растут. Затем ученик снова пробует решить задачу. Занятия длительностью $T_3 \gg \Delta t$ чередуются переменами продолжительностью $T_п \gg \Delta t$.

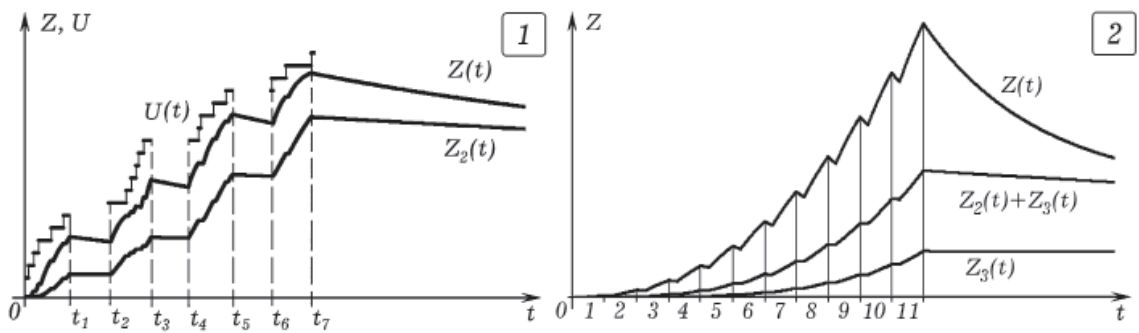

*Рис. 2. Компьютерная имитация процесса обучения:
1 – решение задач возрастающей сложности; 2 – изменение количества знаний при обучении в школе и после ее окончания*

В используемой программе решение задачи рассматривается как случайный процесс, вероятность которого вычисляется по формуле Роша:

$$p_i = 1/(1 + \exp(-\lambda(Z(t) - \theta_i))).$$

При $Z = \theta$ вероятность решения $i$-й задачи равна 0,5. Результаты имитационного моделирования обучения на четырех занятиях представлены на рис. 2.1. Ступенчатая линия $U(t) = \theta(t)$ показывает, как меняется сложность решаемых задач (уровень предъявляемых требований); графики $Z(t) = Z_1 + Z_2$ и $Z_2(t)$ характеризуют динамику роста всех и прочных знаний. Полученные кривые похожи на графики на рис. 1.2, когда требования $U$ в течение урока растут пропорционально времени.

4. Обучение в школе длится 11 лет. Учебный год состоит из 9 месяцев занятий и трех месяцев каникул. Уровень требований, предъявляемых учителем учени-ку в $i$-м классе, задан матрицей ($U_1$, $U_2$, ..., $U_n$). Изучим изменение знаний учащегося во время обучения и после его окончания. Используется трехкомпонентная модель обучения, типичные результаты моделирования представлены на рис. 2.2. Видно, как во время обучения уровень знаний ученика растет, увеличивается количество прочных знаний. Периодические провалы графика $Z(t)$ объясняются забыванием во время каникул. После окончания обучения быстро забываются непрочные знания, которые учащийся редко использовал; прочные знания забываются медленнее.

Трехкомпонентная модель обучения использовалась автором при исследовании проблемы формирования системы эмпирических знаний [6, с. 75–87]. При этом вся совокупность фактов, изучаемых в школе, была разделена на три категории:

1) факты, которые могут быть установлены в повседневной жизни;

2) факты, устанавливаемые в физической лаборатории;

3) факты, не устанавливаемые в условиях обучения и изучаемые умозрительно. После согласования компьютерной модели с результатами педагогического эксперимента были получены графики, характеризующие изменение уровня знаний фактов различных категорий по мере обучения ученика в школе [6, с. 84–85].

### Обобщенная модель обучения

Автором предложена обобщенная модель обучения, не имеющая аналогов в известной ему литературе. Пусть $Z$ – суммарные знания ученика, $Z_1$ – самые непрочные знания первой категории с высоким коэффициентом забывания $\gamma_1$, $Z_2$ – знания второй категории с меньшим коэффициентом забывания $\gamma_2$, ..., а $Z_n$ – самые прочные знания $n$-й категории с низким $\gamma_n$ ($\gamma_1 > \gamma_2 > ... > \gamma_n$)). Коэффициенты усвоения $\alpha_i$ характеризуют быстроту перехода знаний ($i - 1$)-й категории в более прочные знания $i$-й категории. Коэффициент забывания $\gamma = 1/\tau$, где $\tau$ – время, уменьшения знаний в 2,72... раза. Коэффициент сложности $S$ ($0 \leq S \leq 1$) позволяет учесть субъективную сложность усвоения $i$-го ЭУМ.

Обучение характеризуется количеством приобретенных знаний $Z$ и коэффициентом прочности:

$$\Pr = \left( \frac{Z_2}{2^{n-2}} + ... + \frac{Z_{n-1}}{2} + Z_n \right) \bigg/ Z.$$

При изучении одной темы сначала растет уровень знаний $Z$, затем происходит увеличение доли прочных знаний $Z_n$ и повышается прочность Pr. В любой момент времени:

$$Z(t) = Z_1(t) + ... + Z_n(t).$$

Во время обучения:

$$dZ_1/dt = r(1-S)(\alpha_1 F Z^b - \alpha_2 Z_1) - \gamma_1 Z_1;$$
$$dZ_2/dt = r(1-S)(\alpha_2 Z_1 - \alpha_3 Z_2) - \gamma_2 Z_2, ..., dZ_n/dt =$$
$$= r(1-S)\alpha_n Z_{n-1} - \gamma_n Z_n.$$

Время перерыва: $U = 0$, $dZ_1/dt = -\gamma_1 Z_1$, $dZ_2/dt = -\gamma_2 Z_2, ..., dZ_n/dt = -\gamma_n Z_n$.

Использование предложенной модели позволяет проанализировать различные ситуации, встречающиеся в педагогической практике, и учесть влияние сложности изучаемого материала и других факторов на результат обучения [6].

### Заключение

Предлагаемые компьютерные модели дополняют качественные рассуждения, делают их более объективными, обоснованными и могут быть использованы тогда, когда проведение педагогического эксперимента неправомерно или приводит к отрицательным результатам. Изменяя последовательность изучения различных ЭУМ, длительность занятий и т.д., можно найти оптимальный путь обучения в конкретном случае.

Одно из направлений использования имитационного моделирования процесса обучения связано с созданием обучающей программы, моделирующей учебный процесс в школе и предназначенной для тренировки студентов педагогических вузов. Она должна допускать изменение параметров учеников, длительность занятий, распределения учебного материала и стратегии поведения учителя. В процессе ее работы студент, играющий роль учителя, изменяет скорость подачи учебной информации, быстро реагирует на вопросы учеников, проводит контрольные работы, ставит оценки, пытаясь добиться наибольшего уровня знаний за заданное время. После окончания «обучения» на экран выводятся графики, показывающие изменение «количества знаний учеников класса», оценки за

«выполненные контрольные работы» и т.д. Кроме того, обучающая программа анализирует работу «учителя» (студента) и ставит ему оценку.

**Список литературы**